%% file: pap.tex
\documentclass[aps,twocolumn,preprintnumbers,letter]{revtex4}
\usepackage[dvips]{graphicx}
\begin{document}
\title{Static quark anti-quark pair in SU(2) gauge theory}
\author{Alexei Bazavov\email{bazavov@physics.arizona.edu}} \affiliation{
Department of Physics, University of Arizona, Tucson, AZ 85721, USA
}

\author{P\'eter Petreczky\email{petreczk@bnl.gov}} \affiliation{
  RIKEN-BNL Research Center and Physics Department, Brookhaven
  National Laboratory, Upton NY 11973, USA }

\author{Alexander Velytsky \email{vel@uchicago.edu}} 
\affiliation{Enrico Fermi Institute, University of Chicago, 5640 S. Ellis Ave., Chicago, IL 60637, USA,}
\affiliation{HEP Division and Physics Division, Argonne National Laboratory, 9700 Cass Ave., Argonne, IL 60439, USA}

\begin{abstract}

We study singlet and triplet correlation functions of static 
quark anti-quark pair defined through gauge invariant time-like
Wilson loops and Polyakov loop correlators in finite
temperature $SU(2)$ gauge theory. We use the L\"uscher-Weisz
multilevel algorithm, which allows to calculate these correlators
at very low temperatures. We observe that the naive separation
of singlet and triplet states in general does not hold non-perturbatively, 
however, is recovered in the limit of small separation and the 
temperature dependence of the corresponding correlators is
indeed very different.  

\end{abstract} 

\preprint{ANL-HEP-PR-08-45, BNL-NT  08/25}

\maketitle

\section{Introduction}
It is well established that strongly interacting matter undergoes a 
deconfining transition at some temperature which is triggered by large
increase in number of degrees of freedom (e.g. large increase in the
entropy density) as well as melting of hadronic degrees of freedom.
One of the most important feature of the deconfined phase is the screening
of color charges. 
It has been argued that color screening will lead to quarkonium dissociation
above the deconfinement temperature which can be used as a signature of
quark gluon plasma formation in heavy ion collisions \cite{MS86}.
Melting of quarkonium states can be rigorously studied in terms of spectral
functions. Attempts to reconstruct spectral functions from Euclidean time
quarkonium correlators calculated on the lattice have been presented in Refs. 
\cite{umeda02,asakawa04,datta04} (for light mesons see \cite{qcdtaro,karsch02,asakawa03}).
 These studies seemed to indicate  that charmonium states
may survive up to unexpectedly high temperatures $1.6T_c-2.2T_c$ ($T_c$ being
the transition temperature). However, it turns out that reconstruction of
the spectral function is difficult \cite{jako07}. The only statement that can be made
with confidence is that the Euclidean time quarkonium correlation functions
do not show significant temperature dependence \cite{jako07}.

On the lattice color screening is usually studied in terms of the Polyakov
loop correlator related to the free energy of static quark anti-quark pair \cite{mclerran81}.
Unlike quarkonium correlators it shows very significant temperature dependence across the
deconfinement transition. 
In fact, in the deconfined phase the free energy of static quark anti-quark pair shows
large temperature dependence even for very small separations between the static quark and anti-quark,
much smaller than the inverse temperature \cite{okacz00,karsch01,meqm01}. 
In perturbative picture this can be understood due to the fact that in the deconfined phase not only
singlet quark anti-quark ($Q\bar Q$) states contribute to the free energy but also colored states with
$Q\bar Q$ in the adjoint (octet for $SU(3)$ and triplet for $SU(2)$) representation. 
This observation is also supported by lattice
calculations of the correlation function of two temporal Wilson lines in Coulomb gauge, which in 
perturbation theory corresponds to the so-called singlet free energy 
\cite{nadkarni86f1,okacz02,digal03,okacz04,kostya1,okacz05,kostya}. 
The singlet free energy is temperature
independent at short distances and coincides with the zero temperature potential as expected. However, at
larger distances, e.g. distances of the order of typical quarkonium size, it also shows significant temperature dependence.
Thus there seems to be a puzzle: heavy quarkonium correlators show almost no temperature
dependence, while static mesons are largely affected by the deconfined medium.

In the past several years quarkonium properties at finite temperature have been studied
in potential models which use the singlet free energy as an input 
\cite{digal01,wong04,alberico05,mocsy06,rapp07,mocsy07,mocsy07a}. Furthermore, 
within potential model calculations it has been
shown that the small temperature dependence of the quarkonium spectral function does not
necessarily imply survival of quarkonium states at high temperatures \cite{mocsy07,mocsy07a}. 
Most quarkonium
states are melted due to color screening  as it was originally suggested by Matsui and Satz but
threshold enhancement can compensate for absence of bound states \cite{mocsy07,mocsy07a}.

Potential models can be justified in the effective field theory framework, 
the potential non-relativistic QCD (pNRQCD),
where both scales related to the heavy quark mass and 
bound state size are integrated out \cite{brambilla00}.
This approach can be generalized to finite temperature, where also the thermal 
scales (the temperature and the
Debye mass) have to be eventually integrated out \cite{brambilla08}. 
In pNRQCD both color singlet and color octet $Q\bar Q$
states are present as an effective degrees of freedom. 
The problem of defining color singlet and adjoint $Q\bar Q$ states
on the lattice has been considered in Ref. \cite{jahn}. 
It has been found that the conventional definition of singlet and
adjoint states have problems. 
Since this question is very relevant for quarkonium physics at finite temperature as well
as from purely conceptual point of view a more detailed study is needed.

In this
paper we study static meson correlators in 4 dimensional $SU(2)$ gauge theory
at finite temperature and show how the problem observed in 
Ref. \cite{jahn} can be resolved in the limit of small distances and/or
high temperatures. 
The rest of the paper is organized as follows. In section II we discuss static meson correlators
at finite temperature and their interpretation in the limit of high and low temperatures. 
Section III contains
our numerical results along with some important technical details. 
Finally in section IV we give our conclusions.

\section{Static meson correlators}
Following Ref. \cite{jahn} 
we start our discussion considering static meson
operators in color singlet and adjoint (triplet) states
\begin{eqnarray}
&
O(x,y;t)=\bar \psi(x,t) U(x,y;t) \psi(y,t)\\[2mm]
&
O^{\alpha}(x,y;t)=\bar \psi(x,t) U(x,x_0;t) T^{\alpha} U(x_0, y;t) \psi(y,t), 
\end{eqnarray}
where $T^{\alpha}$ is the group generator, $\bar \psi$
and $\psi$ are creation and annihilation operators of static quarks
and $U(x,y;t)$ are the spatial gauge transporters. 
In this section we consider the general case of $SU(N)$ group, although
the numerical calculations have been done for $N=2$.
Next we consider
correlators of these static meson operators at time $t=1/T$ which after integrating
out the static fields have the form \cite{jahn}:
\begin{eqnarray}
&
\displaystyle
G_1(r,T)=\frac{1}{N} 
\langle {\rm Tr} 
L^{\dagger}(x) U(x,y;0) L(y) U^{\dagger}(x,y,1/T) \rangle, \label{defg1}\\[2mm]
&
\displaystyle
G_a(r,T)=\frac{1}{N^2-1}
\langle {\rm Tr} L^{\dagger}(x)  {\rm Tr} L(y) \rangle\nonumber\\[2mm]
&
\displaystyle
-\frac{1}{N (N^2-1)} 
\langle {\rm Tr} L^{\dagger}(x) U(x,y;0) L(y) U^{\dagger}(x,y,1/T) \rangle, \label{defg3}\\
&
r=|x-y|. \nonumber
\end{eqnarray}
Here $L(x)$ is the temporal Wilson line, which on the lattice is simply $L(x)=\prod_{\tau=0}^{N_\tau-1} U_0(x,\tau)$
with $U_0(x,\tau)$ being the temporal links.
The correlators depend on the choice of the spatial transporters
$U(x,y;t)$. Typically a straight line connecting points $x$ and $y$ is
used as a path in the gauge transporters, i.e. one deals with time-like rectangular cyclic
Wilson loops, i.e. Wilson loops wrapping around the time direction.
This object has been calculated at finite temperature in hard thermal loop (HTL) 
petrurbation theory in context of re-summed perturbative 
calculations of quarkonium spectral functions \cite{lpr,laine07,laine08}. 
This is one of the reasons we are interested in non-perturbative evaluation of it.
In the special gauge, where $U(x,y;t)=1$ the above
correlators give standard definition of the so-called singlet and
adjoint free energies  
\begin{eqnarray}
&
\displaystyle
\exp(-F_1(r,T)/T)=
\frac{1}{N} \langle {\rm Tr} L^{\dagger}(x)  L(y)\rangle,\\[2mm]
&
\displaystyle
\exp(-F_a(r,T)/T)=
\frac{1}{N^2-1}
\langle {\rm Tr} L^{\dagger}(x)  {\rm Tr} L(y) \rangle \nonumber\\[2mm]
&
\displaystyle
-\frac{1}{N (N^2-1)} \langle {\rm Tr} L^{\dagger}(x)  L(y)\rangle.
\end{eqnarray}
The singlet and triplet free energies can be calculated at high temperature in
leading order HTL approximation \cite{mehard} resulting in:
\begin{eqnarray}
&
\displaystyle
F_1(r,T)=-\frac{N^2-1}{2 N} \frac{\alpha_s}{r} \exp(-m_D r)-\frac{(N^2-1)\alpha_s m_D }{2 N},
\label{f1p}\nonumber\\
&
\\[2mm]
&
\displaystyle
F_a(r,T)=+\frac{1}{2 N} \frac{\alpha_s}{r} \exp(-m_D r)-\frac{(N^2-1) \alpha_s m_D}{2N},
\label{f3p}
\end{eqnarray}
with $m_D=g T \sqrt{(N/3)} $ being the leading order Debye mass.
At this order $F_1$ and $F_a$ are gauge independent or in other words do not depend
on the choice of
the parallel transporters $U(x,y;t)$.
Note that at small distances ($r m_D\ll 1$) the singlet free energy 
\begin{equation}
F_1(r,T) \simeq -\frac{N^2-1}{2 N} \frac{\alpha_s}{r}
\end{equation}
is temperature independent and coincides with the zero temperature potential, 
while the adjoint free energy 
\begin{equation}
F_a(r,T) \simeq \frac{1}{2 N} \frac{\alpha_s}{r}-\frac{N}{2} \alpha_s m_D 
\end{equation}
depends on the temperature.

The physical free energy of a static 
quark anti-quark pair, i.e. the one related to the work that has to be done to
separate the static charges by certain distance is given by the thermal average
of the singlet and adjoint free energies \cite{mclerran81} 
\begin{eqnarray}
&
\exp(-F(r,T)/T)=\nonumber\\[2mm]
&
\displaystyle
\frac{1}{N^2} \exp(-F_1(r,T)/T) + \frac{N^2-1}{N^2} \exp(-F_a(r,T)/T)\nonumber\\[2mm]
&
\displaystyle
=\frac{1}{N^2} \langle {\rm Tr} L(x) {\rm Tr} L(y) \rangle \equiv \frac{1}{N^2} G(r,T) \label{defg}.
\end{eqnarray}
This quantity is explicitly gauge independent. In leading order HTL approximation
the free energy is 
\begin{equation}
F(r,T)=-\frac{(N^2-1)}{4 N^2} \frac{\alpha_s^2}{r^2 T} \exp(-2 m_D r).
\end{equation}
The $1/r^2$ behavior is due to partial cancellation between the singlet and adjoint contribution
\cite{mclerran81,nadkarni86} and has been confirmed by lattice calculations 
in the intermediate distance regime above deconfinement \cite{meqm01,digal03}.

Using
the transfer matrix one can show that in the confined phase \cite{jahn}
\begin{eqnarray}
&
G_1(r,T)=\sum_{n=1}^{\infty} c_n(r) e^{-E_n(r,T)/T},\label{g1}\\[2mm]
&
G(r,T) = \sum_{n=1}^{\infty} e^{-E_n(r,T)/T},
\label{g}
\end{eqnarray}
where $E_n$ are the energy levels of static quark and anti-quark pair. The 
coefficients $c_n(r)$ depend on the choice of $U$ entering the static meson operator
$O$ in Eq. (1). Since the color averaged correlator $G(r,T)$ corresponds to a measurable quantity (at least in principle)
it does not contain $c_n$. The lowest energy level is the usual static quark anti-quark
potential, while the higher energy levels correspond to hybrid potentials 
\cite{bali01,morningstar99,juge03,michael92}. Using multi-pole  expansion in pNRQCD one can
show that at short distances the hybrid potential corresponds to the adjoint potential
up to non-perturbative constants \cite{brambilla00}. 
Indeed, lattice calculations of the hybrid potentials
indicate a repulsive short distance part \cite{bali01,morningstar99,juge03,michael92}.
If $c_1=1$ the dominant contribution to
$G_a$ would be the first excited state $E_2$, i.e. the lowest hybrid potential which
at short distances is related to the adjoint potential. In this sense $G_a$ is related 
to static mesons with $Q\bar Q$ in adjoint state. 
Numerical calculations show, 
however, that $c_1(r) \ne 1$ and depends on the separation $r$. Thus $G_a$ also
receives contribution from $E_1$ \cite{jahn}. The lattice data seem to suggest that $c_1$ 
approaches unity at short distances \cite{jahn} in accord with expectations based on 
perturbation theory, where $c_1=1$  up to ${\cal O} (g^6)$ corrections  \cite{brambilla00}. 
Therefore  at short distances, $r \ll 1/T$ the color singlet and color averaged free energy
are related $F(r,T)=F_1(r,T) + T \ln (N^2-1)$.
This relation is indeed confirmed by lattice calculations \cite{okacz02}.
In the next section we are going to study
the singlet and averaged correlators $G_1$ and $G$ in  the confined phase and extract
the coefficients $c_1$. 
\input{table_of_runs.tex}

\section{Numerical results}
We have calculated correlation functions of static mesons $G_1(r,T)$ and 
$G(r,T)$ both in  the confined and deconfined phase of $SU(2)$ gauge theory.
In our calculations we used the L\"uscher-Weisz algorithm for noise reduction \cite{luscher}, which
makes possible to calculate  these correlators at low values of temperature $T$ not
accessible by standard methods. Calculations have been done for $\beta=4/g^2=2.5$ and $2.7$ using
lattices with spatial extent $N_{s}=24$ and $N_s=32$. Gauge configurations have 
been generated using combination of heat-bath and over-relaxation algorithms. 
Measurements are taken after
2 complete updates. A complete update consists of one heat-bath and two over-relaxation updates of
the entire lattice and  hundred slice updates, 
in which all links inside a slice $N_s^3\times 2$ except of the boundary
are updated for all $N_\tau/2$ slices with heat-bath.

We have studied the color singlet and averaged correlators
given by Eqs. (\ref{defg1}) and (\ref{defg}). 
The spatial links entering the transporter $U(x,y;0)$ were smeared using APE
smearing \cite{ape}, which has been applied iteratively. The weight of the staple in the APE
smeared link was $0.12$.
For $\beta=2.5$ we use spatial links with 10 steps of
APE smearing and unsmeared spatial links. For $\beta=2.7$
we used unsmeared spatial links as well as spatial links with 10 steps and 20 steps of APE smearing.
The lattice spacing has been set using the string tension calculated in Ref. \cite{michael92}
$a^2 \sigma_{\beta=2.5}=0.0363(3)$ and $a^2 \sigma_{\beta=2.7}=0.0112(2)$. When quoting the
results in terms of reduced  temperature $T/T_c$ we use the value $T_c/\sqrt{\sigma}=0.69(2)$ \cite{fingberg}.
The simulation parameters, including the different levels of APE smearing used in the present
study are summarized in Table \ref{tab:par}.

For comparison with the study of static meson correlation functions in Coulomb gauge \cite{digal03}
we have also performed calculations on $16^3 \times 4$ lattice at $\beta= 2.3533$ and $2.4215$.
The two gauge coupling correspond to temperatures $1.2T_c$ and $1.5T_c$ 
respectively (see Ref. \cite{digal03} for details).
Here we used $5,~10$ and $20$ APE smearings.
The results from these calculations will be discussed in section \ref{sec:deconf_f1}.

\subsection{Color averaged correlator in the confined phase}

\begin{figure}
\includegraphics[width=9.5cm]{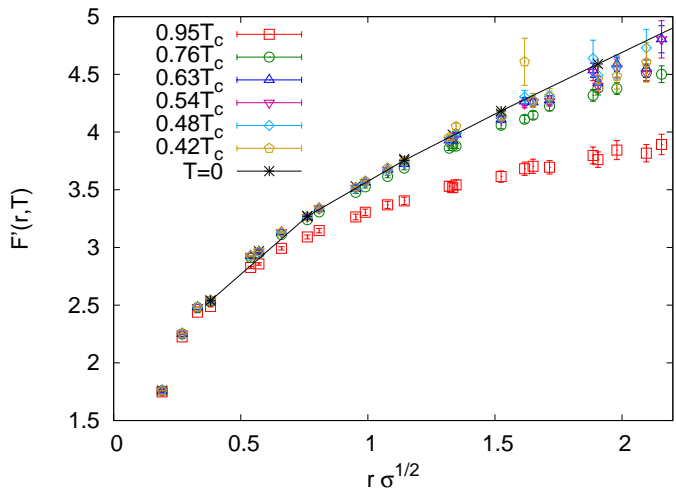}
\caption{The color averaged free energy defined by Eq. (\ref{defg}) below deconfinement
temperature at $\beta=2.5$ calculated on $32^3 \times N_{\tau}$ lattices.
Also shown is the $T=0$ potential.}
\label{fig:fav25}
\end{figure}

The color averaged correlator has been calculated in the
confined phase in the temperature interval $0.32T_c-0.95T_c$ 
$\beta=2.5$ and $0.49T_c-0.98T_c$ for $\beta=2.7$. Thus we study
this correlator also at very low temperatures, where they have not been calculated before.
The numerical results for the color averaged free energy for $\beta=2.5$ are
shown in Figure \ref{fig:fav25}. To eliminate the trivial temperature dependence
due to the color trace normalization in Figure  \ref{fig:fav25} we show the subtracted
free energy $F'(r,T)=F(r,T)-T \ln 4$, see the discussion in the previous section.
In the figure also the zero temperature potential calculated in Ref. \cite{michael92}
is shown. The color averaged free energy does not show any
temperature dependence up to temperatures of about $0.76T_c$.
For $T=0.76T_c$ the color averaged free energy is below the zero temperature potential,
indicating a decrease in the effective string tension in 
accordance with earlier studies (see e.g. \cite{okacz00,digal03}). 
We see even larger change closer to the transition temperature, namely for $T \ge 0.95T_c$.

Since the temperature dependence for $T<0.76T_c$ is relatively small 
we attempted to fit the color
averaged correlator with 
the 1-exponential form $G(r,T)=c_1^a(r) \exp(-E_1(r)/T)$.
The ground state energy $E_1(r)$ extracted from this fit agrees
well with the zero temperature potential calculated in Ref. \cite{michael92},
while the coefficients $c_1^a(r)$ are close to one as expected (see the discussion
in the previous section).
The fit details are shown in the Appendix. Although the deviations of $c_1^a(r)$
from unity are small, they appear to be statistically significant. 
These deviations increase with increasing $r$. Therefore they are likely to be due to the contribution from excited states, 
as the gap between the ground state and excited states (hybrid potential) gets smaller
with increasing separation.
Therefore we attempted to fit the color
averaged correlator with 2-exponential form 
$G(r,T)=\exp(-E_1(r)/T)+\exp(-E_2(r)/T)$ using $T<0.76T_c$.
The results of the fit are given in the Appendix.
We find that  the value of $E_1$ extracted from the 2-exponential fits are consistent
with the ones obtained from 1-exponential fit as well as with the value of the zero
temperature potential. For $E_2(r)$ we find values which are somewhat below the first
hybrid potential. This is presumably due to the fact that there is small contribution
from the higher excited states. 

\subsection{Color singlet correlators in the confined phase}

\begin{figure}
\includegraphics[width=9.5cm]{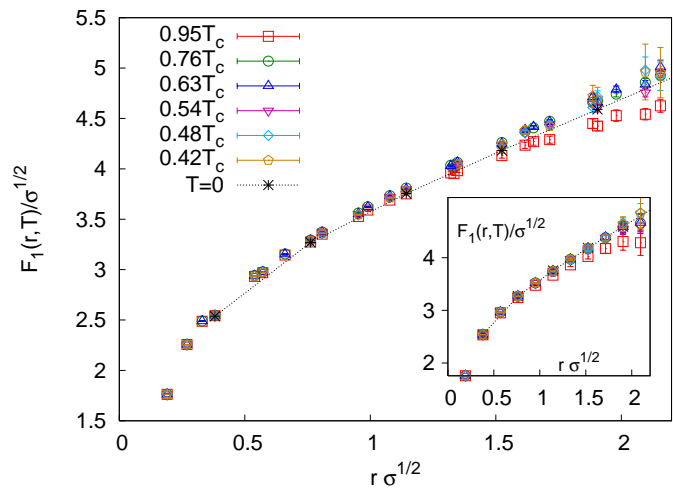}
\caption{The color singlet free energy below deconfinement
temperature at $\beta=2.5$ calculated on $32^3 \times N_{\tau}$ lattices.
Also shown is the $T=0$ potential. The inset shows the color singlet free energy
from which the contribution from the matrix element $T \ln c_1$ has been subtracted. }  
\label{fig:f125}
\end{figure}

\begin{figure}
\includegraphics[width=9.5cm]{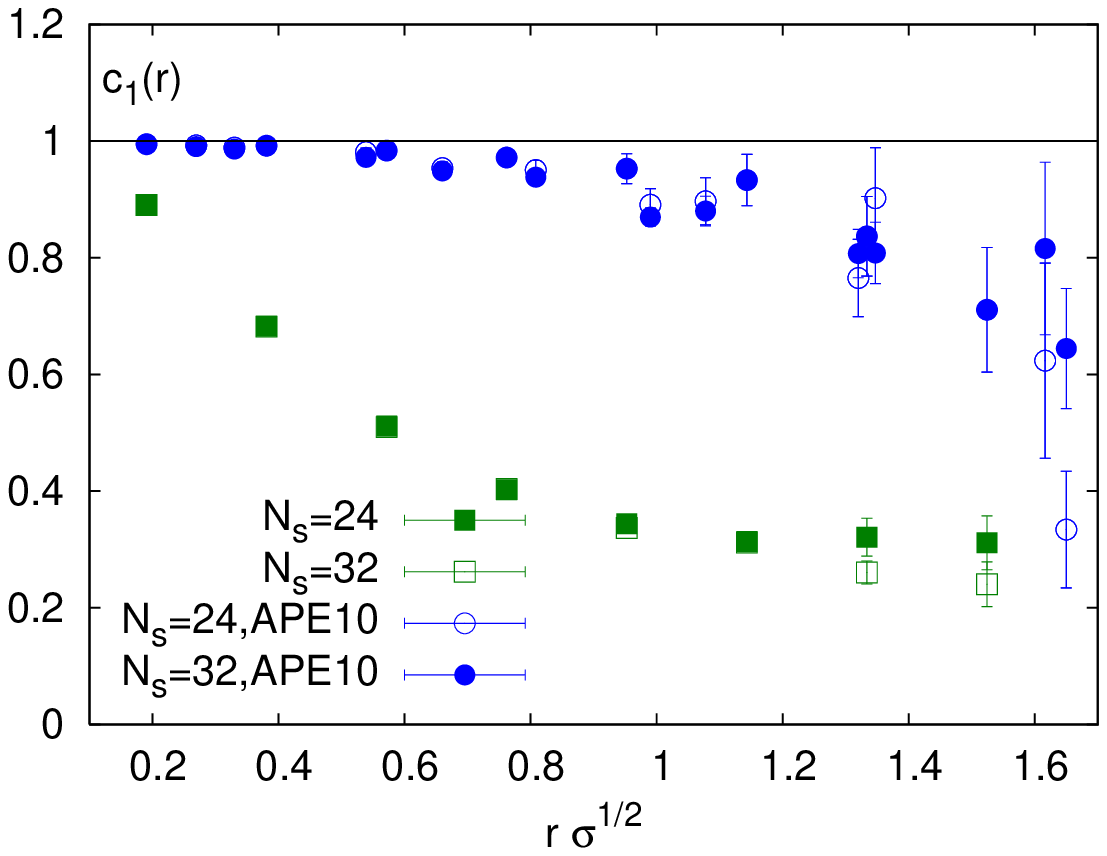}
\includegraphics[width=9.5cm]{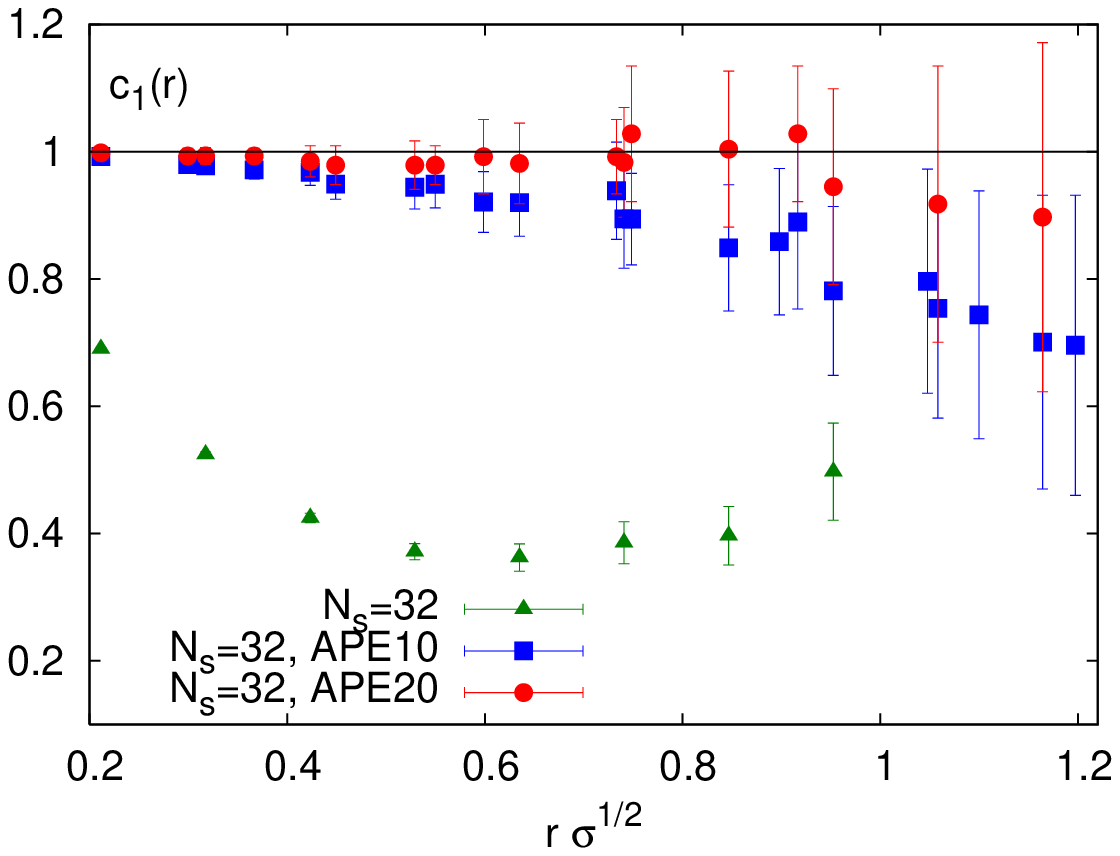}
\caption{The pre-exponential factor of the color singlet correlators as function of
distance $r$ for $\beta=2.5$ (top) and $\beta=2.7$ (bottom). Shown are results
for unsmeared spatial links and 10 and 20 steps of APE smearing.}
\label{fig:c1}
\end{figure}

As mentioned above the  color singlet correlators
have been calculated using different levels of APE smearing in the spatial gauge connection.
It is well known that smearing increases the overlap with the ground state by removing 
the short distance fluctuation in the spatial links (see e.g. \cite{booth}).
For this reason smearing also reduces the breaking of the rotational invariance to the
level expected in the free theory.
We have found that when no smearing is used the color singlet
free energy, $-T \ln G_1(r,T)$ shows a small but visible temperature dependence.
In particular $F_1(r,T)$ is larger than the $T=0$ potential for intermediate
distances $0.5<r \sqrt{\sigma} <2$. A similar  effect has been observed in calculation 
in 3 dimensional $SU(2)$ gauge theory \cite{jahn,ophil02} as well in 4 dimensional $SU(2)$ and $SU(3)$ 
gauge theory calculations in
Coulomb gauge \cite{digal03,okaczlat03}. 
The temperature dependence of the singlet free energy is significantly
reduced when APE smearing is applied.
In Figure \ref{fig:f125} we show the color singlet free energy for $\beta=2.5$
and 10 APE smearings. As one can see from the figure
the color singlet free energy shows significantly smaller temperature dependence
as we get closer to the deconfinement temperature. In particular, only for $T=0.95T_c$
we see  significant temperature dependence, which, however, is much smaller than for
color averaged free energy.
To understand the temperature dependence of the color singlet
correlator we use 1-exponential fit $G_1(r,T)=c_1(r) \exp(-E_1(r)/T)$.
In all cases considered, the value of $E_1(r)$ extracted from fits are
in good agreement with the calculation of the zero temperature potential in Ref.\cite{michael92}.
The value of the prefactor $c_1(r)$ is shown in Figure \ref{fig:c1}.
When no APE smearing is used the value of $c_1(r)$ strongly depends on the separation $r$.
At small distances it shows a tendency of approaching unity as one would expect in perturbation theory.
However, $c_1(r)$ decreases with increasing distance $r$. 
At large distance its value is around $0.3-0.5$ (see Appendix for details).
Similar results for $c_1(r)$ have been obtained in study of $SU(2)$ gauge theory 
in 3 dimensions \cite{jahn}.
When APE smearing is applied the $r$-dependence of the amplitude $c_1(r)$ is largely reduced
and its value is close to unity both for $\beta=2.5$ and $\beta=2.7$. For $\beta=2.7$ we also see
that increasing the number of smearing steps from 10 to 20 reduces the deviation of $c_1(r)$ from unity.

As discussed in section II perturbation theory predicts that the deviations of $c_1(r)$ from unity is
of order $\alpha_s^3$. Therefore it can be made arbitrarily small by going to sufficiently small
distances but even for distances accessible in this study these deviations are expected to be small
based on perturbation theory. It is known, however, that lattice perturbation theory converges very
poorly. The main reason for this has been identified with the short distance fluctuations of link variables,
which makes their mean value very different from unity \cite{lepage92}. Smearing removes these short
distance fluctuations and this is the reason why $c_1(r)$ is much closer to unity when APE smearing is
applied.

From the above discussions it is clear that almost the entire temperature dependence 
of the singlet free energy at distances $0.5<r \sqrt{\sigma}<2$ is due to the deviation 
of $c_1$ from unity and can be largely reduced by applying APE smearing to
the links in the spatial gauge connections. To further demonstrate this point in the inset of
Figure \ref{fig:f125} we show the results for $F_1(r,T)+T \ln c_1(r)$. Clearly no temperature dependence
can be seen in this quantity up to $0.95T_c$, where we see temperature dependence at 
distances $r\sqrt{\sigma} \ge 1.5$ 
corresponding the expected drop of the 
effective string tension.

\subsection{Color singlet free energy in the deconfined phase}
\label{sec:deconf_f1}
In this subsection we discuss the properties of the color singlet free energy,
$F_1(r,T)=-T \ln G_1(r,T)$ above the deconfinement temperature. It turns out that
the singlet free energy calculated from cyclic Wilson loops shares the same qualitative 
features as the singlet free energy calculated
in Coulomb gauge \cite{okacz02,digal03,okacz04,kostya1,okacz05}. At short distances it is
temperature independent and coincides with the zero temperature potential.
At large distances it approaches a constant $F_{\infty}(T)$ which monotonically
decreases with the temperature. The constant $F_{\infty}(T)$  is the free energy of 
two isolated static quarks, or equivalently of a quark anti-quark pair at infinite separation. 
Its value is 
therefore independent of the definition 
of the singlet correlator $G_1(r,T)$ and is related to the renormalized
Polyakov loop $L_{ren}(T)=\exp(-F_{\infty}(T)/(2 T))$ \cite{okacz02}. 

\begin{figure}
\includegraphics[width=9.4cm]{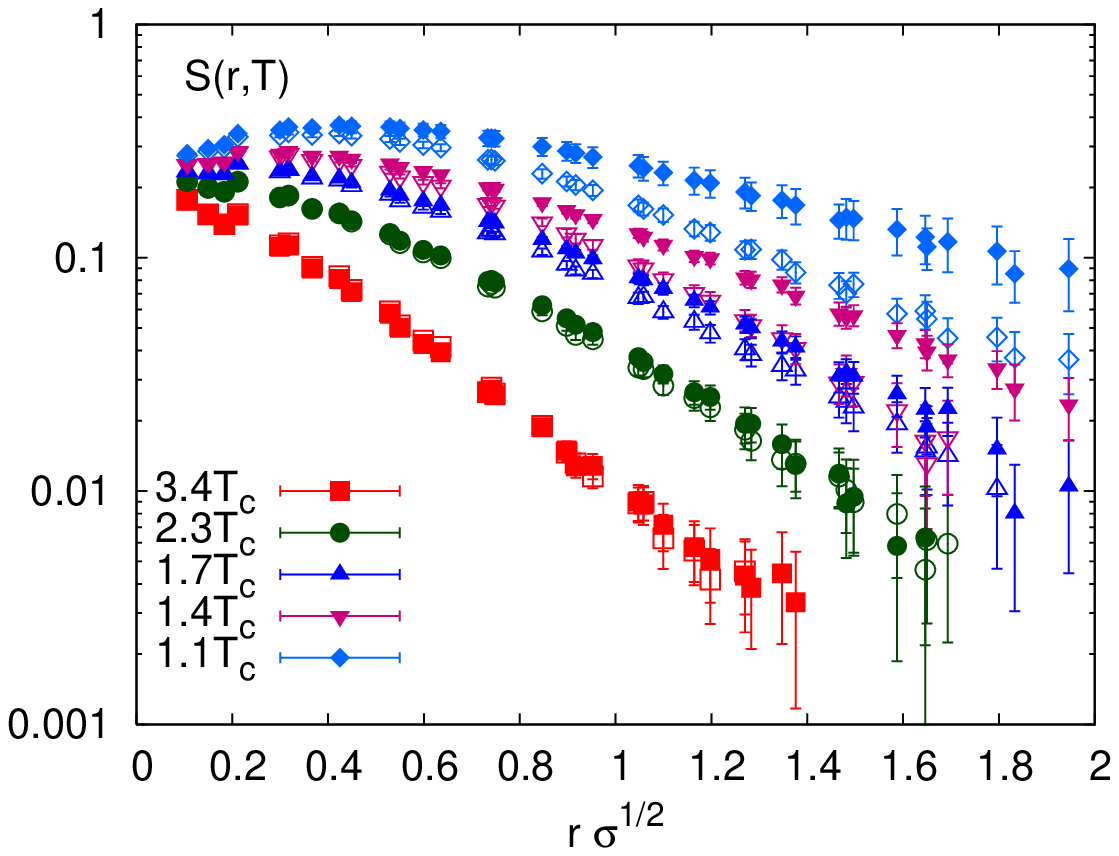}
\includegraphics[width=9.4cm]{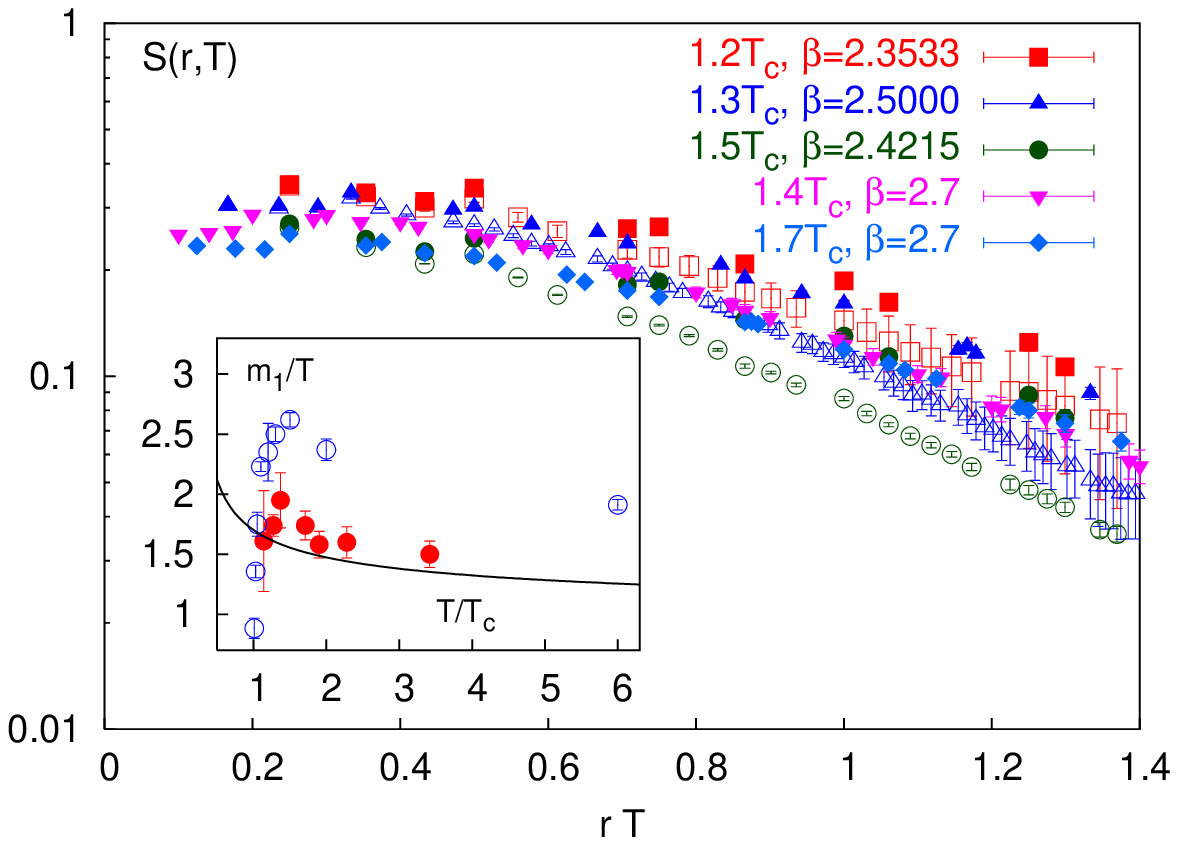}
\caption{The screening function $S(r,T)=r (  F_1(r,T)-F_{\infty}(T))$ at different
temperatures calculated for $\beta=2.7$ as function of the distance in units of $\sqrt{\sigma}$ (top) and 
as function of $r T$ and different values of $\beta$ (bottom). 
In the lower panel we also shown the results from calculations in 
Coulomb gauge \cite{digal03} as open symbols.
In the inset the screening masses $m_1$ extracted
from singlet free energies are shown together with the results obtained
in Coulomb gauge \cite{digal03} (open symbols).
The line shows the leading order results for the Debye mass.
In the upper panel the filled symbols  correspond to
20 APE smearings steps, while the open ones to 10 APE smearings steps.} 
\label{fig:s1}
\end{figure}

At leading order  $F_1(r,T)-F_{\infty}(T)$ is of Yukawa form (c.f. Eq. (\ref{f1p})). Therefore
we find it useful to show our numerical results in terms of the screening function 
\begin{equation}
S(r,T)=r \cdot (  F_1(r,T)-F_{\infty}(T))
\end{equation}
In Figure \ref{fig:s1} we show the results for the screening function at different temperatures.
At short distances ($r T<0.5$) the singlet free energy 
does not depend on the smearing level. Furthermore, it is very close to the free energy
calculated in Coulomb gauge. 
We expect that at large distances the screening function $S(r,T)$ will show an exponential 
decay determined by a temperature dependent screening mass $m_1(T)$, which is equal to the leading order
Debye mass up to the non-perturbative $g^2$ corrections: $m_1=m_D + {\cal O}(g^2)$ \cite{rebhan}.
From Fig. \ref{fig:s1}   we can see  that indeed $S(r,T)$ behaves exponentially  with screening mass
proportional to the temperature. We note, however,
that 
there is some dependence on the smearing level at larger distances. This disappears
at high temperatures and with increasing the smearing level. In particular, for $\beta \le 2.5$
it turns out that there is no dependence on smearing level for 5 or more smearing steps.
For $\beta=2.7$ we need 10-20 steps depending on the temperature to achieve stable results.
Fitting the large distance behavior of the screening function by an exponential form $\exp(-m_1(T) r)$ we
determine the screening mass $m_1(T)$.
Typically we considered distances $r >1/T$ as well as the maximal number of smearings 
steps (10 for $\beta \le 2.5$ and 
20 for $\beta=2.7$ ) in our fits.
In the inset of Fig. \ref{fig:s1} we  show the color singlet screening masses 
extracted from the fits and compare them
to the results obtained in Coulomb gauge in Ref. \cite{digal03}.
In the inset we also show the leading order Debye mass calculated using 2-loop gauge
coupling $g(\mu=2 \pi T)$ in $\overline{MS}$-scheme. 
We used the value $T_c/\Lambda_{\overline{MS}}=1.09$ \cite{rank97} to calculate the coupling $g$.
As we see from the figure the screening
masses are smaller than those calculated in Coulomb gauge and agree well with the leading order
perturbative prediction. 

Leading order perturbation theory also predicts that
$F_1^2(r,T)/F(r,T)=6$ for $r>1/T$ (c.f. Eqs. \ref{f1p}-\ref{f3p}). We find that at the highest
temperature, $T=3.42T_c$ this relation   indeed holds within statistical errors.

\subsection{Color triplet free energy}
We have calculated the color triplet correlator defined by Eq. (\ref{defg3})
for different temperatures below and above the transition temperature.
Below the deconfinement temperature we observe a moderate $T$-dependence of the triplet
correlator. We also find that the corresponding free energy $-T\ln G_3(r,T)$ 
is smaller than the first hybrid potential calculated in Ref. \cite{michael92}, but larger than the triplet
free energy  in Coulomb gauge \cite{digal03}. 

\begin{figure}
\includegraphics[width=8cm]{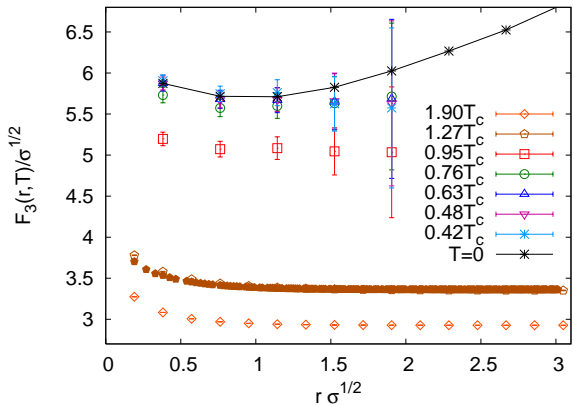}
\caption{The triplet free energy at different temperatures calculated
at $\beta=2.5$. 
The filled symbols correspond to calculations in Coulomb gauge.
Also shown is the first hybrid potential calculated in Ref. \cite{michael92}.
}
\label{fig:f3corr}
\end{figure}

Let us assume that only two states contribute to the Eqs. (\ref{g1}) and (\ref{g}).
Then from Eq. (\ref{defg3}) it follows that
\begin{equation}
F_3(r,T)=E_2(r)-T\ln(1-c_2(r)+\frac{1}{3}(1-c_1(r)) e^{\Delta E(r)/T}),
\label{f3_2comp}
\end{equation}
with $\Delta E(r)=E_2(r)-E_1(r)$. We have seen in section III.B that the temperature dependence of the 
singlet free energy is quite small. In any case it is considerably smaller than the temperature
dependence of the averaged free energy. Therefore the contribution of the excited states
to $G_1(r,T)$ is quite small and it is reasonable to assume that $c_2(r) \ll 1$.
We also expect that at small distances, $c_2(r) \sim (r \Lambda_{QCD})^4$ \cite{antonio}. 
Thus, the temperature dependence of $F_3(r,T)$ and its deviation from the hybrid state $E_2(r)$ is
due to small deviation of $c_1(r)$ from unity. At low temperatures, when $\Delta E \gg T$ these small
deviations are amplified by the exponential factor.
To verify this, we have subtracted the correction $T \ln (1+\frac13(1-c_1) e^{\Delta E/T})$ from the triplet
free energy assuming that $E_1(r)$ is given by the ground state potential and $E_2(r)$ is given by the
first hybrid potential as calculated in Ref. \cite{michael92}.
The numerical results are summarized in Fig. \ref{fig:f3corr} which shows that after this 
correction is accounted for in the confined phase the triplet
free energy at low temperatures agrees reasonably well with the first hybrid potential. As temperature
increases more excited states contribute. In particular,  at $0.76T_c$ the value of the triplet
free energy can be accounted for by including the next hybrid state \cite{michael92}.
However, at  $0.95T_c$ we see large temperature effects, which cannot be explained by including the contribution from 
only few excited states.

In Fig. \ref{fig:f3corr} we also show the triplet free energy above the deconfinement temperature
compared to the calculations in Coulomb gauge \cite{digal03}. It turns out to be much smaller than in the
confined phase and agrees well with Coulomb gauge results. This means that the small deviation of
the overlap factor $c_1(r)$ from unity are unimportant in this case. The triplet free energy monotonically
decreases with increasing temperature as expected in HTL perturbation theory (c.f. Eq. (\ref{f3p})). 
In the limit of high temperatures and short distances $r<1/T$ we have 
$E_2(r)=\alpha_s/(4 r),~\Delta E(r)=\alpha_s/r,c_2(r) \simeq 0$ and $c_1(r)=1+{\cal O}(\alpha_s^3)$.
Therefore we can expand the logarithm in Eq. (\ref{f3_2comp}) to get 
\begin{equation}
F_3(r,T)=+\frac{1}{4} \frac{\alpha_s}{r}+{\cal O}(\alpha_s^3 T)+{\cal O}(\alpha_s m_D).
\end{equation}
Thus the correction due to $c_1(r) \ne 0$ are much smaller than the expected leading order
thermal effects in the triplet free energy.

While the temperature dependence of $F_3(r,T)$ and $F_1(r,T)$ is different even at short
distances the $r$-dependence is expected to be similar. 
In leading order perturbation theory we have 
$$
\frac{\displaystyle \frac{d F_1(r,T)}{d r}}{\displaystyle \frac{d F_3(r,T)}{d r}}=-3.
$$
We have calculated the above ratio numerically from the lattice data and have found that
for $T>1.71T_c$ the deviation from the expected value ($-3$) are smaller than $10\%$.

\section{Conclusions}

In this paper we have studied singlet and triplet static quark anti-quark correlators 
in finite temperature $SU(2)$ theory expressed in terms  of Polyakov loop correlators 
and cyclic Wilson loops (cf. Eqs (\ref{defg1}), (\ref{defg3})). 
The study of the latter is interesting as it has been used in resummed perturbative
calculations of quarkonium spectral functions \cite{lpr,laine07}.
In leading order and probably next-to-leading order of perturbation theory
the static correlators defined by Eq. (\ref{defg1}) and (\ref{defg3}) give the energies of the 
singlet and triplet states respectively, however, this separation does not
hold in general case. Due to interactions with ultrasoft fields there will be a mixing
of singlet and triplet states  which is proportional to $\alpha_s^3(1/r)$ and $(r \mu)^4$,
with $\mu$ being the ultra-soft scale \cite{brambilla00}. 
In our case the ultra-soft scale can be the  binding energy, $\alpha_s/r$, $~\Lambda_{QCD}$ or $~g^2 T$.
Therefore it is expected that mixing is quite small at sufficiently small distances.
We determined the mixing between singlet and triplet states in terms of the overlap
factor $c_1(r)$. If the overlap factor is unity there is no mixing. Our lattice
calculations show that $c_1(r)$ indeed approaches one at small distances. Using
iterative APE smearing the deviation of $c_1(r)$ from unity can be largely reduced. 
Therefore the contribution of singlet state to $G_3(r,T)$ appears to be small at
temperatures close to deconfinement temperature. This contribution is also
controlled by the non-perturbative gap between the singlet and triplet states, i.e.
the gap between the static potential and the first hybrid potential.

Our analysis shows that at short distances $r T <1 $ the singlet
correlator is almost temperature independent, while the  
triplet correlator is largely affected by the deconfinement.
The temperature dependence of the triplet correlators indicate the melting of the
non-perturbative gap between the singlet and the triplet states above deconfinement,
which turns out to be consistent with perturbative expectations. Because of the
disappearance of the non-perturbative gap the small deviations of $c_1(r)$ from unity
play no role in the deconfined phase. In particular, the difference between the
singlet and triplet correlators defined through cyclic Wilson loops and using
Coulomb gauge turns out to be quite small. This finding is important for application
of thermal pNRQCD discussed in Ref. \cite{brambilla08} to realistic quarkonia and
temperatures not very far from the deconfinement temperature.

There are obvious extensions of the work carried out in this paper. 
One could do analogous calculations in $SU(3)$ gauge theory, however this
hardly will give qualitatively different results. More interesting would
be to extend the calculations to QCD with dynamical fermions. First steps
in this direction are being made  using improved staggered (p4)
action  \cite{future}. It will be interesting to calculate the correlators
$G_1(r,T)$ and $G_3(r,T)$ in perturbation theory beyond HTL approximation.
One practical problem in doing these calculations is the fact that the scales $T$ and
$m_D$ are not separated. This problem, however, could be possibly solved by using
screened perturbation theory \cite{scpt97,braaten,blaizot}.

\begin{acknowledgments}
This work was supported by U.S. Department of Energy under
Contract No. DE-AC02-98CH10886.
The work of A.B. was supported by grants DOE DE-FC02-06ER-41439 and NSF 0555397.
A.V. work was supported by the Joint Theory Institute funded together by
Argonne National Laboratory and the University of Chicago, 
and in part by the U.S. Department of Energy, 
Division of High Energy Physics and Office of Nuclear Physics, under Contract DE-AC02-06CH11357.
We thank N. Brambilla, F. Karsch D. Teaney and A. Vairo for useful discussions.
\end{acknowledgments}

\section*{Appendix}
\input{appendix.tex}

\end{document}

%% file: table_of_runs.tex


\begin{table}[hp]
\centering
\begin{tabular}{|c|c|c|c|c|c|c|c|c|c|c|}
\hline
\multicolumn{5}{|c|}{$\beta=2.5$} & &  \multicolumn{5}{c|}{$\beta=2.7$} \\
\hline
$N_s$ & $N_\tau$ & $T/T_c$ & $N_{APE}$ & \# meas & &
$N_s$ & $N_\tau$ & $T/T_c$ & $N_{APE}$ & \# meas  \\ \hline
 32   &     4    &   1.90  &    0      & 4K    & &
 32   &     4    &   3.42  &    0      & 4K     \\ \hline
 32   &     4    &   1.90  &   10      & 1K    & &
 32   &     4    &   3.42  &   10, 20  & 1K     \\ \hline
 32   &     6    &   1.27  &    0      & 4K    & &
 32   &     6    &   2.28  &    0      & 4K     \\ \hline
 24,32&     6    &   1.27  &   10      & 1K,1K & &
 32   &     6    &   2.28  &   10, 20  & 1K     \\ \hline
 24,32&     8    &   0.95  &    0      & 16K,4K & &
 32   &     8    &   1.71  &    0      & 4K     \\ \hline
 24,32&     8    &   0.95  &   10      & 1K,1K & &
 32   &     8    &   1.71  &   10, 20  & 1K     \\ \hline
 24,32&    10    &   0.76  &    0      & 8K,4K & &
 32   &    10    &   1.37  &    0      & 8K     \\ \hline
 24,32&    10    &   0.76  &   10      & 1K,1K & &
 32   &    10    &   1.37  &   10, 20  & 1K     \\ \hline
 24,32&    12    &   0.63  &    0      & 8K,4K & &
 32   &    12    &   1.14  &    0      & 8K  \\ \hline
 24,32&    12    &   0.63  &   10      & 1K,1K & &
 32   &    12    &   1.14  &   10, 20  & 1K     \\ \hline
 24,32&    14    &   0.54  &    0      & 4K,4K & &
 32   &    14    &   0.98  &    0      & 8K     \\ \hline
 24,32&    14    &   0.54  &   10      & 1K,1K  & &
 32   &    14    &   0.98  &   10, 20  & 1K     \\ \hline
 24,32&    16    &   0.48  &    0      & 4K,4K & &
 32   &    16    &   0.86  &    0      & 4K     \\ \hline
 24,32&    16    &   0.48  &   10      & 1K,1K & &
 32   &    16    &   0.86  &   10, 20  & 1K     \\ \hline
 24,32&    18    &   0.42  &    0      & 8K,4K & &
 32   &    18    &   0.76  &    0      & 8K     \\ \hline
 24,32&    18    &   0.42  &   10      & 1K,1K & &
 32   &    18    &   0.76  &   10, 20  & 1K     \\ \hline
 24,32&    20    &   0.38  &    0      & 8K,4K & &
 32   &    20    &   0.68  &    0      & 4K     \\ \hline
 24,32&    20    &   0.38  &   10      & 1K,1K & &
 32   &    20    &   0.68  &   10, 20  & 1K     \\ \hline
 24,32&    22    &   0.35  &    0      & 8K,4K & &
 32   &    22    &   0.62  &    0      & 4K     \\ \hline
 24,32&    22    &   0.35  &   10      & 1K,1K & &
 32   &    22    &   0.62  &   10, 20  & 1K     \\ \hline
 32   &    24    &   0.32  &    0      & 4K    & &
 32   &    24    &   0.57  &    0      & 4K     \\ \hline
 32   &    24    &   0.32  &   10      & 1K    & &
 32   &    24    &   0.57  &   10, 20  & 1K     \\ \hline
      &          &         &           &       & &
 32   &    28    &   0.49  &    0      & 4K     \\ \hline
      &          &         &           &       & &
 32   &    28    &   0.49  &   10, 20  & 1K     \\ \hline
\hline
\end{tabular}
\caption{Simulation parameters, \# meas refers to the
actual number of measurements. 
}
\label{tab:par}
\end{table}


%% file: appendix.tex
In this appendix we give some details of the fits of the overlap factors 
and energy levels for different lattice volumes.

Let us first discuss the fits of the color averaged correlator $G(r,T)$.
We have fitted our lattice data at low temperatures with one exponential form
$G(r,T)=c_1^a(r)\exp(-E_1(r)/T)$.
For $\beta=2.5$ we used lattices with temporal extent 
$N_{\tau}=12,~14,~16,~18,~20,~22$ and $24$. For $\beta=2.7$ we used
$N_{\tau}=20,22,24$ and $28$. The values of fit parameters $c_1^a(r)$ and
$E_1(r)$ are shown
in Table \ref{tab:g_1exp_25} for $\beta=2.5$ and Table \ref{tab:g_1exp_27}
for $\beta=2.7$. 
As one can see from the Tables at $\beta=2.5$ the single exponential fit
works reasonably well up to distance $r=7$ for $N_s=32$ and distance $r=6$
for $N_s=24$. At $\beta=2.7$ the fits work up to distance $r=5$ only.
We also performed double exponential fits 
$G(r,T)=\exp(-E_1(r)/T)+\exp(-E_2(r)/T)$ to check the reliability of the
extraction of $E_1$. The results are shown in Tables \ref{tab:g_2exp_25}
and \ref{tab:g_2exp_27} for $\beta=2.5$ and $\beta=2.7$ respectively.
Overall the values of $E_1$ extracted from 2-exponential fits are in reasonably
good agreement with the corresponding values extracted from singlet exponential
fit.

\begin{table}
\begin{tabular}{|r|l|l|l|l|l|l|}
\hline
\multicolumn{1}{|c|}{} &  \multicolumn{3}{|c|}{$N_s=24$}  & \multicolumn{3}{|c|}{$N_s=32$} \\
\hline
$r$ & $c_1^a$     & $E_1$        & $\frac{\chi^2}{\mbox{dof}}$
   & $c_1^a$     & $E_1$        & $\frac{\chi^2}{\mbox{dof}}$ \\ \hline
 1 & 1.00077(28) & 0.33524(2)   & 0.50   & 1.00046(28) & 0.33522(1) & 1.48 \\
 2 & 1.0049(13)  & 0.48423(8)   & 0.49   & 1.0039(12)  & 0.48415(6) & 0.96 \\
 3 & 1.0134(45)  & 0.56568(24)  & 0.64   & 1.0115(32)  & 0.56550(16)  & 0.93 \\
 4 & 1.027(10)   & 0.62400(61)  & 0.73   & 1.0237(67)  & 0.62365(35)  & 0.75 \\
 5 & 1.049(22)   & 0.6732(14)   & 1.23   & 1.038(14)   & 0.67232(78)  & 0.66 \\
 6 & 1.067(43)   & 0.7171(27)   & 1.43   & 1.051(29)   & 0.7161(17)   & 1.10 \\
 7 & 1.087(91)   & 0.7582(59)   & 1.80   & 1.023(60)   & 0.7542(40)   & 1.53 \\
 8 & 1.05(16)    & 0.793(11)    & 2.06   & 0.93(11)    & 0.7861(83)   & 1.69 \\
 \hline
\end{tabular}
\caption{Single exponential fit of $G(r,T)$  at
$\beta=2.5$ using $N_{\tau}=12-24$ for two spatial sizes $N_s=24$ and $N_s=32$.}
\label{tab:g_1exp_25}
\end{table}

\begin{table}
\begin{tabular}{|r|l|l|l|}
\hline
 $r$ &   $c_1^a$        & $E_1$          & $\frac{\chi^2}{\mbox{dof}}$ \\
\hline
 1 &   1.00065(44)    & 0.28580(2)     & 0.64 \\
 2 &   1.0084(28)     & 0.39584(12)    & 1.21 \\
 3 &   1.0325(90)     & 0.44727(36)    & 0.90 \\
 4 &   1.078(21)      & 0.47939(80)    & 0.54 \\
 5 &   1.143(39)      & 0.5038(14)     & 0.41 \\
\hline
\end{tabular}
\caption{Single exponential fit of $G(r,T)$ at
$\beta=2.7$ using $N_t=20-28$.}
\label{tab:g_1exp_27}
\end{table}

\begin{table}
\begin{tabular}{|r|l|l|l|l|l|l|}
\hline
\multicolumn{1}{|c|}{} & \multicolumn{3}{|c|}{$N_s=24$} & \multicolumn{3}{|c|}{$N_s=32$} \\
\hline
$r$ & $E_1$        & $E_2$       & $\frac{\chi^2}{\mbox{dof}}$ &
    $E_1$        & $E_2$       & $\frac{\chi^2}{\mbox{dof}}$ \\
 \hline
1 & 0.335204(5)  & 1.011(41)   & 0.89 & 0.335199(3)  & 1.048(50) & 0.84 \\
2 & 0.483974(18) & 0.989(20)   & 0.56 & 0.483948(9)  & 0.998(15) & 0.47 \\
3 & 0.565022(42) & 0.983(17)   & 0.44 & 0.564946(29) & 0.982(14) & 0.60 \\
4 & 0.62269(12)  & 0.984(21)   & 0.55 & 0.622522(88) & 0.986(19) & 0.80 \\
5 & 0.67088(42)  & 0.987(36)   & 1.08 & 0.67046(19)  & 1.003(24) & 0.65 \\
6 & 0.7140(10)   & 1.004(57)   & 1.35 & 0.71357(48)  & 1.023(40) & 0.86 \\
7 & 0.7545(24)   & 1.016(86)   & 1.69 & 0.7533(12)   & 1.07(10)  & 1.43 \\
\hline
\end{tabular}
\caption{Double exponential fit of $G(r,T)$  at
$\beta=2.5$  using $N_{\tau}=12-24$ for two spatial sizes $N_s=24$ and $N_s=32$. }
\label{tab:g_2exp_25}
\end{table}

\begin{table}
\begin{tabular}{|r|l|l|l|}
\hline
 $r$ & $E_1$        &  $E_2$       & $\frac{\chi^2}{\mbox{dof}}$ \\
\hline
 1 &  0.285780(1)  &  0.700(11)   &    0.14 \\
 2 &  0.395551(9)  &  0.6819(60)  &    0.15 \\
 3 &  0.446179(27) &  0.6657(45)  &    0.11 \\
 4 &  0.476943(53) &  0.6532(33)  &    0.07 \\
 5 &  0.49965(12)  &  0.6445(37)  &    0.07 \\
 6 &  0.51857(29)  &  0.6362(48)  &    0.11 \\
 7 &  0.53545(67)  &  0.6297(63)  &    0.16 \\
 8 &  0.5514(14)   &  0.6240(79)  &    0.18 \\
\hline
\end{tabular}
\caption{Double exponential  fit of $G(r,T)$ at
$\beta=2.7$ using $N_{\tau}=20-28$.}
\label{tab:g_2exp_27}
\end{table}

\begin{table}
\begin{tabular}{|r|l|l|l|l|l|l|}   
\hline
\multicolumn{1}{|c|}{} & \multicolumn{3}{|c|}{$N_s=24$} & \multicolumn{3}{|c|}{$N_s=32$} \\
\hline
$r$ & $c_1$     & $E_1$        & $\frac{\chi^2}{\mbox{dof}}$ 
    & $c_1$     & $E_1$        & $\frac{\chi^2}{\mbox{dof}}$ \\ 
\hline
 1 & 0.89076(34)   & 0.33523(2)  & 0.36 & 0.89043(28) & 0.33521(2)  & 1.31 \\
 2 & 0.68239(89)   & 0.48406(8)  & 0.77 & 0.68191(79) & 0.48401(6)  & 0.70 \\
 3 & 0.5113(16)    & 0.56549(20) & 0.20 & 0.5095(14)  & 0.56519(14) & 1.11 \\
 4 & 0.4020(33)    & 0.62353(53) & 0.19 & 0.4035(24)  & 0.62360(33) & 0.58 \\
 5 & 0.3439(63)    & 0.6731(12)  & 0.82 & 0.3367(46)  & 0.67123(86) & 0.34 \\
 6 & 0.313(14)     & 0.7174(32)  & 2.24 & 0.312(10)   & 0.7173(21)  & 1.32 \\
 7 & 0.321(32)     & 0.7642(74)  & 1.93 & 0.261(20)   & 0.7496(54)  & 1.10 \\
 8 & 0.311(46)     & 0.803(11)   & 0.70 & 0.240(38)   & 0.786(11)   & 3.19 \\
 9 & 0.35(13)      & 0.845(29)   & 1.43 & 0.228(79)   & 0.822(24)   & 1.69 \\
\hline
\end{tabular}
\caption{Single exponential fit of $G_1(r,T)$ at different lattice separations $r$ at
$\beta=2.5$ with no APE smearing using  $N_{\tau}=12-24$.}
\label{tab:g1_ape00_25}
\end{table}

\begin{table}
\begin{tabular}{|r|l|l|l|l|l|l|}   
\hline
\multicolumn{1}{|c|}{} & \multicolumn{3}{|c|}{$N_s=24$} & \multicolumn{3}{|c|}{$N_s=32$} \\
\hline
$r$ & $c_1$     & $E_1$        & $\frac{\chi^2}{\mbox{dof}}$  
& $c_1$     & $E_1$        & $\frac{\chi^2}{\mbox{dof}}$ \\
\hline
1.0000 & 0.99457(96) & 0.33528(6)  & 0.71 & 0.99260(50) & 0.33516(3)  & 0.53 \\
1.4142 & 0.9926(19)  & 0.42930(11) & 0.81 & 0.9904(11)  & 0.42916(6)  & 0.32 \\
1.7320 & 0.9886(26)  & 0.47322(16) & 0.38 & 0.9864(16)  & 0.47310(9)  & 0.33 \\
2.0000 & 0.9919(30)  & 0.48414(17) & 0.70 & 0.9882(18)  & 0.48389(10) & 0.19 \\
2.8284 & 0.9810(69)  & 0.55824(42) & 0.53 & 0.9721(40)  & 0.55758(24) & 0.32 \\
3.0000 & 0.9835(63)  & 0.56554(36) & 0.50 & 0.9751(46)  & 0.56494(27) & 0.28 \\
3.4641 & 0.954(10)   & 0.59681(68) & 0.72 & 0.9487(60)  & 0.59643(37) & 0.47 \\
4.0000 & 0.972(13)   & 0.62406(82) & 0.33 & 0.9503(98)  & 0.62243(60) & 0.19 \\
4.2426 & 0.950(16)   & 0.6376(11)  & 0.33 & 0.938(11)   & 0.63656(70) & 0.32 \\
5.0000 & 0.953(26)   & 0.6732(17)  & 0.83 & 0.923(18)   & 0.6706(12)  & 0.40 \\
5.1961 & 0.891(28)   & 0.6803(21)  & 1.47 & 0.870(16)   & 0.6787(11)  & 0.72 \\
5.6568 & 0.897(40)   & 0.7015(32)  & 0.94 & 0.880(25)   & 0.7000(18)  & 0.12 \\
6.0000 & 0.933(44)   & 0.7185(31)  & 0.48 & 0.877(35)   & 0.7129(26)  & 0.54 \\
6.9282 & 0.765(66)   & 0.7473(61)  & 1.34 & 0.807(41)   & 0.7508(33)  & 0.88 \\
7.0000 & 0.837(68)   & 0.7550(61)  & 0.62 & 0.874(64)   & 0.7563(49)  & 0.38 \\
7.0711 & 0.902(87)   & 0.7643(71)  & 0.99 & 0.808(52)   & 0.7568(44)  & 0.29 \\
8.0000 & 0.71(11)    & 0.786(11)   & 1.28 & 0.86(10)    & 0.7963(80)  & 1.78 \\
\hline
\end{tabular}
\caption{Single exponential fit of $G_1(r,T)$ at 
$\beta=2.5$ calculated with 10 steps of APE smearing using $N_{\tau}=12-24$.}
\label{tab:g1_ape10_25}
\end{table}

\begin{table}
\begin{tabular}{|r|l|l|l|}   
\hline
 $r$ & $c_1$     & $E_1$        & $\frac{\chi^2}{\mbox{dof}}$ \\ 
\hline
   1  &    0.89048(47)  &   0.28578(2)  &  2.72\\
   2  &    0.6898(18)   &   0.39557(11) &  2.30 \\
   3  &    0.5241(39)   &   0.44628(31) &  2.29 \\ 
   4  &    0.4243(73)   &   0.47792(72) &  1.66 \\
   5  &    0.371(12)    &   0.5019(14)  &  1.44 \\
   6  &    0.362(21)    &   0.5238(25)  &  0.83  \\ 
   7  &    0.385(33)    &   0.5449(36)  &  0.36 \\
   8  &    0.396(46)    &   0.5612(48)  &  1.04 \\
   9  &    0.497(76)    &   0.5838(64)  &  0.74 \\
  10  &    0.55(11)     &   0.5991(89)  &  0.59 \\ 
\hline
\end{tabular}
\caption{Single exponential fit of $G_1(r,T)$ at
$\beta=2.7$ with no APE smearing using $N_{\tau}=20-28$.}
\label{tab:g1_ape00_27}
\end{table}

\begin{table}
\begin{tabular}{|r|l|l|l|}   
\hline
 $r$ & $c_1$     & $E_1$        & $\frac{\chi^2}{\mbox{dof}}$ \\ 
\hline
   1.0000   &   0.9962(12)  &    0.28582(5)  & 0.02   \\ 
   1.4142   &   0.9952(25)  &    0.35819(10) & 1.00   \\
   1.7320   &   0.9927(33)  &    0.38969(14) & 0.59   \\
   2.0000   &   0.9923(42)  &    0.39585(18) & 0.18   \\
   2.8284   &   0.9798(97)  &    0.44311(42) & 1.74   \\
   3.0000   &   0.978(11)   &    0.44681(45) & 0.59   \\
   3.4641   &   0.971(14)   &    0.46541(62) & 1.19   \\
   4.0000   &   0.968(20)   &    0.47854(88) & 0.41   \\
   4.2426   &   0.949(24)   &    0.4852(11)  & 1.08   \\ 
   5.0000   &   0.944(34)   &    0.5018(15)  & 0.22   \\
   5.1961   &   0.949(37)   &    0.5073(17)  & 0.89   \\
   5.6568   &   0.921(48)   &    0.5151(22)  & 1.02   \\
   6.0000   &   0.920(52)   &    0.5211(24)  & 0.04   \\ 
   6.9282   &   0.939(76)   &    0.5398(34)  & 0.78   \\
   7.0000   &   0.894(78)   &    0.5381(37)  & 0.20   \\   
   7.0711   &   0.894(72)   &    0.5399(34)  & 0.60   \\
   8.0000   &   0.849(99)   &    0.5522(50)  & 0.02   \\
   8.4853   &   0.86(11)    &    0.5614(57)  & 0.30   \\
   8.6603    &  0.89(14)    & 0.5661(64)    &  0.37 \\
   9.0000   &   0.78(13)    &    0.5640(73)  & 0.24   \\
   9.8995   &   0.80(18)    &    0.5793(94)  & 0.07   \\
  10.0000   &   0.75(17)    &    0.5767(98)  & 0.19   \\
  10.3923   &   0.74(19)    & 0.584(11)      &  0.25 \\
\hline
\end{tabular}
\caption{Single exponential fit of $G_1(r,T)$ at 
$\beta=2.7$ calculated with 10 steps of APE smearing using $N_{\tau}=20-28$.}
\label{tab:g1_ape10_27}
\end{table}

\begin{table}
\begin{tabular}{|r|l|l|l|}   
\hline
   $r$ & $c_1$     & $E_1$         & $\chi^2/\mbox{dof}$ \\ 
\hline
   1.0000  &  0.9965(15)  & 0.28582(6)    &  1.21 \\
   1.4142  &  0.9965(26)  & 0.35809(11)   &  1.10 \\
   1.7320  &  0.9957(45)  & 0.38949(19)   &  2.02 \\
   2.0000  &  0.9982(53)  & 0.39573(22)   &  1.29 \\
   2.8284  &  0.993(12)   & 0.44262(51)   &  1.05 \\
   3.0000  &  0.993(13)   & 0.44639(55)   &  1.38 \\
   3.4641  &  0.985(21)   & 0.46418(88)   &  3.86 \\
   4.0000  &  0.985(24)   & 0.4771(10)    &  2.50 \\
   4.2426  &  0.979(31)   & 0.4839(13)    &  0.65 \\
   5.0000  &  0.979(38)   & 0.4999(16)    &  2.64 \\
   5.1962  &  0.984(48)   & 0.5049(20)    &  3.10 \\
   5.6569  &  0.992(58)   & 0.5139(25)    &  0.86 \\
   6.0000  &  0.982(63)   & 0.5193(27)    &  2.34 \\
   6.9282  &  1.04(10)    & 0.5378(40)    &  2.29 \\
   7.0000  &  0.983(86)   & 0.5360(37)    &  2.10 \\
   7.0711  &  1.03(11)    & 0.5395(44)    &  0.86 \\
   8.0000  &  1.00(12)    & 0.5525(52)    &  2.15 \\
   8.4853  &  1.11(18)    & 0.5640(70)    &  1.08 \\
   9.0000  &  0.94(15)    & 0.5642(69)    &  1.67 \\
  10.0000  &  0.92(22)    & 0.577(10)     &  1.82 \\
\hline
\end{tabular}
\caption{Single exponential fit of $G_1(r,T)$  at 
$\beta=2.7$ calculated for 20 APE smearing using $N_{\tau}=20-28$.}
\label{tab:g1_ape20_27}
\end{table}

We fitted the singlet correlators calculated for different
number of APE smearing steps with single exponential Ansatz
$G_1(r,T)=c_1(r) \exp(-E_1(r)/T)$. The results for the fit parameters
$c_1(r)$ and $E_1(r)$ at $\beta=2.5$ are shown in Tables \ref{tab:g1_ape00_25} and
\ref{tab:g1_ape10_25} for zero and 10 APE smearing steps respectively.
In Tables \ref{tab:g1_ape00_27}, \ref{tab:g1_ape10_27} and \ref{tab:g1_ape20_27}
we show the fit parameters for $\beta=2.7$ and zero, $10$ and $20$ APE smearing steps.